\documentclass[12pt]{article}
\usepackage{amsmath,amssymb}
\usepackage[dvips]{graphicx}

\begin{document}

\begin{titlepage}
\begin{flushright}
    OU-HET 462  \\
    hep-th/0401026  \\
    January 2004
\end{flushright}
\begin{center}
  \vspace{3cm}
  {\bf \Large Gravitational Dielectric Effect and Myers Effect}
  \\  \vspace{2cm}
  Yoshifumi Hyakutake\footnote{E-mail :hyaku@het.phys.sci.osaka-u.ac.jp}
   \\ \vspace{1cm}
   {\it Department of Physics, Osaka University, 
   Toyonaka, Osaka 560-0043, Japan}
\end{center}

\vspace{2cm}
\begin{abstract}
   
In this paper we study the gravitational dielectric phenomena
of a D2-brane in the background of Kaluza-Klein monopoles and D6-branes.
In both cases the spherical D2-brane with nonzero radius becomes classical solution
of the D2-brane action.
We also investigate the gravitational Myers effect in the background of
D6-branes. This phenomenon occurs since the tension of the D2-brane
balances with the repulsive force between D0-branes and D6-branes.

\end{abstract}
\end{titlepage}

\setlength{\baselineskip}{0.65cm}

\vspace{1cm}
\section{Introduction}

In the electromagnetism, oppositely charged particles under 
the uniform electric field separate from each other 
to cancel out the background flux. This is the well-known 
dielectric effect, and analogous phenomena can be observed in
the type II superstring theories and the M-theory\cite{Mye1}.

Let us illustrate the dielectric phenomenon in the type IIA superstring
theory with the case where a spherical D2-brane is located in the 
background of coincident NS5-branes\cite{BDS}. 
NS5-branes carry the magnetic charge of the NS-NS 2-form potential, 
and its 3-from flux vertically penetrates three dimensional sphere
$S^3$ which encloses NS5-branes in the transverse space.
The D2-brane is assumed to be spherical in the $S^3$.
In case there is no magnetic flux on the D2-brane, 
it collapses because of its tension.
On the other hand, if the magnetic flux exists on the D2-brane, 
the background NS-NS 3-form flux supports it against the 
collapse\footnote{Here we fix the position of the D2-brane
in the radial direction in the transverse space 
of NS5-branes. When a fluctuation around the radial direction is taken into account,
the spherical D2-brane falls in the throat of NS5-branes\cite{Hya2,Hya3}.
The final state will be a giant graviton in the M-theory\cite{MST,DJM,Asa}.}.
Notice that the number of the magnetic flux on the D2-brane 
is less than that of NS5-branes.
This is the dielectric effect in the type IIA superstring theory\footnote{
Dielectric effect in flux branes background is considered in ref.~\cite{CHC}.}.

Interestingly the dielectric phenomena in the type II superstring theories
admit the dual descriptions. Namely similar dielectric effects are observed
from the dynamics of D0-branes, which are identical to the magnetic
flux on the D2-brane\cite{ML,Dou}. The dynamical variables of $M$ coincident D0-branes 
are $U(M)$ gauge field and nine adjoint scalars, and the latter represent the position
of D0-branes when they are simultaneously diagonalized.
In the background of NS5-branes, D0-branes form a fuzzy (noncommutative) geometry
in the $S^3$ because of the support from the NS-NS 3-form flux\cite{ARS,ARS2,HNS}.
In fact D0-branes compose a fuzzy sphere in the $S^3$, which matches
the D2-brane picture when the number of D0-branes or magnetic flux is sufficiently large.
This is called the Myers effect\cite{Mye2}.

Since the essence of the dielectric effect and the Myers effect is
the existence of the background flux, at first sight it seems impossible to
observe these effects in purely gravitational backgrounds.
Note, however, that the NS5-brane is related via T-duality to 
a Kaluza-Klein monopole which forms the purely gravitational geometry\cite{GP,OV}.
Therefore it is natural to expect gravitational
versions of the dielectric effect and the Myers effect in the background of 
Kaluza-Klein monopoles\footnote{The instability of coincident D0-branes
under the backgrounds of hyperboloid and Schwarzschild black hole is argued
in ref. \cite{BG}.}.

In this paper, we examine the spherical D2-brane in the background of
Kaluza-Klein monopoles in the type IIA superstring theory.
The geometry of the 10 dimensional Kaluza-Klein monopoles is described by 
$\mathbb{R}^{1,5} \times M_4$, where the $M_4$ is equivalent to 
$\mathbb{R}^3 \times S^1$ but the $S^1$ vanishes at the origin. 
The D2-brane is spherically wrapped around the origin of the $\mathbb{R}^3$
and has momentum along the $S^1$. Then we will see that the spherical D2-brane 
with some finite radius becomes the classical solution of the D2-brane action. 
Since the background is purely gravitational,
this is exactly the case where the dielectric effect is induced via the gravity.

The above system is easily reinterpreted in the framework of the 
11 dimensional M-theory.
The 10 dimensional Kaluza-Klein monopole is just lifted to a 11 dimensional 
Kaluza-Klein monopole, $\mathbb{R}^{1,5} \times M_4 \times S^1$, 
and the D2-brane is done to a M2-brane.
Of course the M2-brane moving along the $S^1$ in the $M_4$ is spherically stable 
around the origin of the $\mathbb{R}^3$. This is the gravitational dielectric effect
in the M-theory.

Now we change the roles of the two $S^1$s in the above discussion, which is often called
the 9-11 flip in the M-theory.
After the dimensional reduction to the 10 dimensional spacetime,
we obtain spherical D2-D0 bound state in the background of D6-branes in the
type IIA superstring theory\cite{Tow}. As discussed before, this system will admit the dual
description of D0-branes, that is, the Myers effect.
We will see that the fuzzy sphere composed by D0-branes really becomes a classical 
solution of equations of motion in the background of D6-branes.
This is the gravitational Myers effect in the type IIA superstring theory.

The organization of this paper is as follows.
In section 2, we discuss the gravitational dielectric effect in the background
of Kaluza-Klein monopoles and D6-branes in the type IIA superstring theory.
In section 3, we observe the gravitational Myers effect in the background
of D6-branes. The stability of the spherical configuration of the D2-D0 bound state
in the background of D6-branes is examined in section 4.
Section 5 is devoted to conclusions and discussions.

\section{The Gravitational Dielectric Effects}

\subsection{The dielectric effect in the background of Kaluza-Klein monopoles}

In this section we study the dielectric effects which are induced
not by the background flux but by the gravitational force.
At first sight it seems difficult to observe the gravitational dielectric 
phenomena in the absence of the background flux.
However, as mentioned in the introduction, the spherical D2-D0 bound state
is stable in the background of NS5-branes, and by the T-duality transformation
this background is related to the Kaluza-Klein monopoles
which are purely gravitational objects. 
Therefore it is interesting to study whether the gravitational 
dielectric effect can be observed in the background of Kaluza-Klein monopoles.

In the following we investigate a single D2-brane with $M$ units of magnetic 
flux on its world-volume in the background of $N$ coincident Kaluza-Klein monopoles.
The Kaluza-Klein monopoles are purely gravitational objects
and the metric for the $N$ coincident Kaluza-Klein monopoles is given by\cite{GP}
\begin{alignat}{3}
  &ds^2 = \eta_{\mu\nu} dx^\mu dx^\nu + f(dr^2 + r^2 d\theta^2 
  + r^2 \sin^2 \theta d\phi^2) + f^{-1} R^2 ( d\psi - A_{\phi} d\phi )^2,
  \label{eq:KKmono}
  \\
  &f = 1 + \frac{N R}{2r}, \quad 
  A_{\phi} = - \frac{N}{2} \cos \theta, \quad
  0 \le \theta \le \pi, \quad 0 \le \phi,\psi \le 2\pi, \notag
\end{alignat}
where $\mu,\nu = 0,\cdots,5$ and $R$ is the radius of the $\psi$
direction at $r = \infty$. When the above solution is reduced to $(r,\theta,\phi)$ 
space, it describes $N$ coincident magnetic monopoles which are localized at $r=0$.

Dynamical degrees of freedom on the D2-brane are $U(1)$ gauge field and 
scalar fields, which represent the position of the D2-brane.
And the effective action for the D2-brane is described by the Born-Infeld action
and the Chern-Simons action:
\begin{alignat}{3}
  S_{\text{D}2} &= -T_2 \int d^3\xi e^{-\phi} \sqrt{-\det 
  \big(P[G]_{ab} + \lambda F_{ab} \big)}
  + \mu_2 \int \big( C^{(3)} + C^{(1)}\wedge \lambda F \big). \label{eq:actD2}
\end{alignat}
Here $T_2$ and $\mu_2$ are tension and charge of a D2-brane, respectively.
These are exspressed as $T_2 = \mu_2 = \frac{1}{(2\pi)^2\ell_s^3 g_s}$
by using the string length $\ell_s$ and the string coupling $g_s$.
The notation $P[\cdots]$ represents the pullback on the world-volume of the D2-brane
and $\xi^a (a=0,1,2)$ are the world-volume coordinates.
$F_{ab}$ is the $U(1)$ gauge field strength on the D2-brane and $\lambda$ is
defined as $\lambda = 2\pi\ell_s^2$.
We assume that the D2-brane is spherical and moving along the $\psi$ direction 
in the background of Kaluza-Klein monopoles.
So the world-volume coordinates $\xi^a$ are identified with $(t=x^0,\theta,\phi)$, and
$d\psi = \dot{\psi}dt$.
Now the D2-brane is located in the background of Kaluza-Klein monopoles
which are purely gravitational objects, the relevant part of the D2-brane action 
is only the Born-Infeld part.

It is straightforward to evaluate Born-Infeld part of the D2-brane action.
By substituting these assumptions for the action, it becomes as
\begin{alignat}{3}
  S_{\text{D}2} 
  &= - T_2 \int dt d\theta d\phi f r^2 
  \sqrt{\sin^2\theta + \Big(\frac{R A_\phi}{fr}\Big)^2
  - f^{-1} R^2 \dot{\psi}^2 \sin^2\theta } .
\end{alignat}
In order to examine the stability of this system, it is convenient to
switch to the Hamiltonian formalism. The canonical momentum conjugate to
the variable $\psi$ is given by
\begin{alignat}{3}
  P &= T_2 f r^2 \frac{f^{-1} R^2 \dot{\psi} \sin^2\theta }
  {\sqrt{\sin^2\theta + (\frac{R A_\phi}{fr})^2
  - f^{-1} R^2 \dot{\psi}^2 \sin^2\theta }} . 
\end{alignat}
Note that the $\psi$ direction is compactified with the circle,
so the total momentum should be conserved as
\begin{alignat}{3}
  M = \int d\theta d\phi P, \label{eq:conmom}
\end{alignat}
where $M$ is an integer.
Then we obtain the Hamiltonian of the form
\begin{alignat}{3}
  \mathcal{H} 
  &= \int d\theta d\phi T_2 f r^2 \frac{\sin^2\theta + (\frac{R A_\phi}{fr})^2}
  {\sqrt{\sin^2\theta + (\frac{R A_\phi}{fr})^2
  - f^{-1} R^2 \dot{\psi}^2 \sin^2\theta }} \label{eq:E1}
  \\
  &= \int d\theta d\phi
  \sqrt{\sin^2\theta + \Big(\frac{R A_\phi}{fr}\Big)^2}
  \sqrt{ \big(T_2 fr^2\big)^2 + f \Big(\frac{P}{R \sin\theta}\Big)^2 }. \label{eq:E2}
\end{alignat}
The interpretation of this Hamiltonian (\ref{eq:E2}) is as follows.
The first square root part is the correction to the volume factor of the 
sphere $\sin\theta$, because of the nontrivial winding of $\phi$ along 
the compactified direction. Terms in the second square root are the sum
of the mass squared of the spherical D2-brane
and the momentum energy squared along compactified direction.
Notice that the radii of sphere and compactified direction are 
affected by gravitational factor $f$, respectively.
These modifications are expected from the form of the metric (\ref{eq:KKmono}).

Let us examine the stability of the spherical D2-brane in the background
of Kaluza-Klein monopoles. In the region of $NR \ll 2r$, 
the factor $f$ is approximated as $f \sim 1$, and the spherical D2-brane shrinks to
a point because of its tension. On the other hand, near the region of $r = 0$,
the spherical D2-brane tends to expand because of the repulsive force between
the momentum along the $\psi$ direction and the Kaluza-Klein monopoles.
Thus we obtain the gravitational dielectric effect.

In order to see this effect in detail, we focus on the neighborhood of the 
Kaluza-Klein monopoles.
In this region, $2r \ll NR$, the factor $f$ is approximated as
$f \sim \tfrac{NR}{2r}$, and the conservations of the momentum (\ref{eq:conmom}) 
and the energy (\ref{eq:E1}) are written like
\begin{alignat}{3}
  M &= 4\pi T_2 R f^{\tfrac{1}{2}} r^2 a^{-1} \big(K(a^2) - E(a^2)\big),
  \\
  \mathcal{H} &= 4\pi T_2 fr^2 K(a^2),
\end{alignat}
where $a(r) \equiv f^{-\frac{1}{2}}R \dot{\psi}$. We also introduced $K(a^2)$ and $E(a^2)$
which are the complete elliptic integrals of the first kind and second kind, respectively. 
By combining these equations, we see that the energy is given 
as a function of the radius $r$. The energy takes the minimum with finite
redius $r=r_\ast$. (See fig.~\ref{fig:D6}.)
Note that the analysis here is reliable if the parameters satisfy the
condition of $\ell_s \ll 2 r_\ast \ll NR$, and this holds actually when $N$ and $M$
are chosen as sufficiently large.

\subsection{The dielectric effect in the background of D6-branes} \label{sec:D6}

In the previous subsection, the gravitational dielectric phenomenon of the spherical D2-brane
with the momentum in the background of Kaluza-Klein monopoles is observed 
in the type IIA superstring theory.
This system is easily lifted to the 11 dimensional M-theory, and 
we obtain a dielectric effect of a spherical M2-brane
with momentum along the compactified direction $\psi$
in the background of 11 dimensional Kaluza-Klein monopoles.

Again we reduce the 11 dimensional spacetime to the 10 dimensional one, 
by identifying the $\psi$ direction with the reduced 10th one.
Then the 11 dimensional Kaluza-Klein monopoles become 
D6-branes in the type IIA superstring theory\cite{Tow}.
And the M2-brane with the momentum along the reduced $\psi$ direction
turns into a bound state of a D2-brane and D0-branes.
Consequently the dielectric effect of the spherical D2-D0 bound state
in the background of D6-branes will be obtained in the type IIA superstring theory.
In this subsection we will see this phenomenon from the viewpoint of the effective theory on
the D2-brane, where D0-branes are identical to the magnetic flux on it.

The D6-brane is extending to $(1+6)$ dimensional spacetime and
carries the magnetic charge of the R-R 1-form potential $C^{(1)}$.
The classical solution of $N$ coincident D6-branes is given by\cite{GM}
\begin{alignat}{3}
  &ds^2 = f^{-\frac{1}{2}} \eta_{\mu\nu}dx^\mu dx^\nu 
  + f^{\frac{1}{2}} (dr^2 + r^2 d\theta^2 + r^2 \sin^2\theta d\phi^2), 
  \\
  &e^\phi = f^{-\frac{3}{4}} , \quad f = 1 + \frac{N \ell_s g_s}{2r} , \quad
  G^{(2)} = \frac{N\ell_s g_s}{2} \sin\theta d\theta \wedge d\phi, \notag
\end{alignat}
where $\mu,\nu = 0,\cdots,6$ and 
$G^{(2)}$ is the R-R 2-form flux originating from the
D6-branes.

As in the previous subsection, the D2-brane is embedded into the background 
of D6-branes spherically, so the world-volume coordinates on the D2-brane
$\xi^a$ are identified with $(t=x^0,\theta,\phi)$.
On the D2-brane we introduce electric field $E=F_{t\theta}$ and magnetic field $B=F_{\theta\phi}$.
Total magnetic flux on the spherical D2-brane is given by 
$M = \frac{1}{2\pi} \int d\theta d\phi B$.

\begin{figure}[tb]
\begin{center}
  \includegraphics[width=9cm,height=4cm,keepaspectratio]{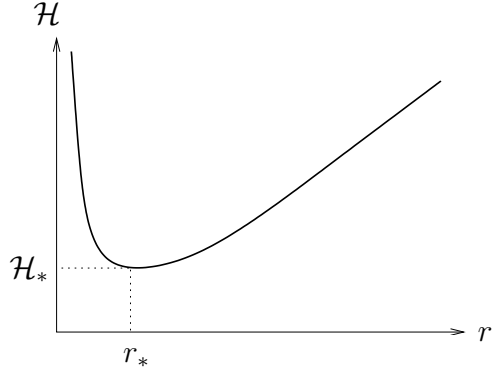}
\begin{picture}(300,0)
  \put(66,132){$\mathcal{H}$}
  \put(234,12){$r$}
  \put(100,4){$r_\ast$}
  \put(57,36){$\mathcal{H}_\ast$}
\end{picture}
  \vspace{0cm}
  \caption{The energy of the spherical D2-brane.}
  \label{fig:D6}
\end{center}
\end{figure}

Now it is straightforward to evaluate the D2-brane action,
and the Lagrangian density is given by
\begin{alignat}{3}
  \mathcal{L} &= -T_2 \sqrt{f^2r^4 \sin^2\theta + f \lambda^2 B^2
  - f^2 r^2 \lambda^2 E^2 \sin^2\theta} - \frac{N}{4\pi}E \cos\theta. \label{eq:D2lag}
\end{alignat}
The momentum conjugate to $A_\theta$ is defined as
\begin{alignat}{3}
  \Pi = \frac{T_2f^2 r^2 \lambda^2 E \sin^2\theta}{\sqrt{f^2r^4 \sin^2\theta + f \lambda^2 B^2
  - f^2 r^2 \lambda^2 E^2 \sin^2\theta}} - \frac{N}{4\pi}\cos\theta,
\end{alignat}
and we obtain the Hamiltonian of the form
\begin{alignat}{3}
  \mathcal{H} = \int d\theta d\phi \sqrt{ \sin^2\theta + 
  \Big( \frac{\Pi + \frac{N}{4\pi}\cos\theta}{T_2 \lambda fr} \Big)^2 }
  \sqrt{ \big(T_2 fr^2\big)^2 + f \Big( \frac{T_0 B}{2\pi \sin\theta} \Big)^2}. \label{eq:pot}
\end{alignat}
Terms in the second square root are mass squared of the spherical D2-brane
and $M$ D0-branes, respectively.
Both terms are modified by the gravitational factor $f$, as compared 
with the flat case. (See fig. \ref{fig:D6}.)

Let us examine the stability of the spherical D2-brane in the background
of D6-branes. Far from the D6-branes the factor $f$ is approximated as 
$f \sim 1$, and the spherical D2-brane shrinks to reduce the total energy. 
On the other hand, near the D6-branes the factor $f$ is approximated 
as $f \sim \tfrac{N\ell_s g_s}{2r}$, and the D2-brane tends to expand to reduce the energy.
Therefore the spherical D2-brane with some finite radius becomes the classical solution.

Now we concentrate on the neighborhood of the D6-branes, where the factor
$f$ is described as $f \sim \tfrac{N\ell_s g_s}{2r}$.
In order to estimate the energy of the D2-brane, we employ the equations of
motion for gauge fields, which say variables $\Pi$ and
$\delta \mathcal{L}/\delta (\frac{B}{2\pi})$ only depend on the radius.
So we choose solutions as $\Pi = 0$ and $\delta \mathcal{L}/\delta (\frac{B}{2\pi}) = 
f^{\frac{1}{2}}(\ell_s g_s)^{-1} a(r)$, and the magnetic field is expressed as
\begin{alignat}{3}
  \frac{B}{2\pi} &= (2\pi\lambda)^{-1} f^{\frac{1}{2}} r^2 
  \frac{a\sin^2\theta}{\sqrt{1-a^2\sin^2\theta}}. \label{eq:D2mag}
\end{alignat}
Then total magnetic flux and the energy are given by
\begin{alignat}{3}
  M &= 2 \lambda^{-1} f^{\tfrac{1}{2}} r^2 a^{-1} \big(K(a^2) - E(a^2)\big),
  \\
  \mathcal{H} &=\int d\theta d\phi \sqrt{ \big(T_2 fr^2\big)^2 
  + f \Big( \frac{T_0 B}{2\pi \sin\theta} \Big)^2} = 4\pi T_2 fr^2 K(a^2). \label{eq:D2ene}
\end{alignat}
The energy becomes a function of the radius, 
and it takes minimum $\mathcal{H}_\ast$ with finite radius $r_\ast$. (See fig.~\ref{fig:D6}.)
This is the gravitational dielectric effect of the spherical D2-D0 bound state 
in the background of D6-branes.
The analyses in this paragraph are reliable if the parameters satisfy 
the condition of $\ell_s \ll 2 r_\ast \ll N\ell_s g_s$, and
this is satisfied by choosing $N$ and $M$ as sufficiently large.
It is worth mentioning that above relations are obtained
by setting $R=\ell_s g_s$, $\dot{\psi} = \delta \mathcal{L}/\delta (\frac{B}{2\pi})$ 
and $P=B/2\pi$ in the previous subsection.

\section{The Gravitational Myers Effect}

D0-branes become dynamical through the excitations of open strings ending on them.
And the massless excitations correspond to the $U(M)$ gauge field $A_0$ and
nine adjoint scalars $X^m (m=1,\cdots,9)$, where $M$ is the number of the D0-branes\cite{Wit}.
When three of nine adjoint scalars do not commute simultaneously,
D0-branes form a fuzzy surface. In particular, D0-branes compose
a fuzzy sphere when three adjoint scalars satisfy the $SU(2)$
Lie algebra. In the background of flat spacetime, the fuzzy sphere is unstable 
and shrinks to a point because of its tension.

On the other hand, as discussed in the introduction, 
the fuzzy sphere becomes stable in the background
of NS5-branes, because the tension balances with the dielectric force 
induced by the background NS-NS 3-form flux.
Furthermore the fuzzy spherical configuration of D0-branes is also stable
in the background of constant R-R 4-form flux.
In both cases the background flux plays an important role to lead to the Myers effect.

In this section we consider another type of Myers effect which is induced 
via the gravitational force.
In the previous section we observed the gravitational dielectric phenomena
in the background of Kaluza-Klein monopoles and D6-branes.
Since the dielectric effect admits the dual description, i.e., the Myers 
effect, the gravitational Myers effect will also be observed in these
backgrounds. 

The metric and R-R 1-form potential for $N$ coincident D6-branes are written as before.
Here it is convenient to employ not polar coordinates but rectangular coordinates like
\begin{alignat}{3}
  &ds^2 = f^{-\frac{1}{2}} \eta_{\mu\nu}dx^\mu dx^\nu 
  + f^{\frac{1}{2}} \delta_{ij}dx^i dx^j, \quad
  e^\phi = f^{-\frac{3}{4}} , \quad f = \frac{N\ell_s g_s}{2r} , \notag
  \\
  &C^{(1)} = -\frac{N\ell_s g_s}{2r} \frac{x^9}{(x^7)^2 + (x^8)^2} (x^7 dx^8 - x^8 dx^7)
  \label{eq:D6-2}
\end{alignat}
where $\mu,\nu = 0,\cdots,6$, which label the tangent directions to D6-branes
and $i,j=7,8,9$, which describe normal ones.
The radius $r$ is given by $r = \sqrt{\delta_{ij}x^i x^j}$,
and the constant term in $f$ is neglected because we only consider the region
near the D6-branes.

The bosonic part of the effective action for $M$ D0-branes is described by the 
nonabelian Born-Infeld action and Chern-Simons action~\cite{Tse,Tse2,TR,Mye2}. 
As explained in the above, this action possesses $U(M)$ gauge symmetry
and consists of $U(M)$ gauge field $A_0$ and nine adjoint scalars 
$X^m (m=1,\cdots,9)$ which appear from massless excitations of open strings
ending on D0-branes. The explicit form of the action 
for $M$ D0-branes is written as
\begin{alignat}{3}
  &\mathcal{S}_{\text{D}0} = - T_0 \int dt \, \text{STr} 
  \big( e^{-\phi} \sqrt{-P[E+E']_{00} \det {Q^m}_n} \big)
  + \mu_0 \int dt \, \text{STr} \big( C^{(1)}_i D_0 X^i \big),
\end{alignat}
where $T_0$ is the mass of a D0-brane.
Note that STr represents the symmetrized trace prescription, and 
partial derivatives in the pullbacks should be replaced 
with covariant derivatives to ensure the $U(M)$ gauge symmetry.
The field $E_{MN}$ and $M \times M$ matrices $E'_{MN}$ and ${Q^m}_n$ are 
defined as
\begin{alignat}{3}
  &E_{MN} = G_{MN} + B_{MN}, \quad
  E'_{MN} = E_{Mm} \big( {(Q^{-1})^m}_n - {\delta^m}_n \big) E^{nl} E_{lN},
  \\
  &{Q^m}_n = {\delta^m}_n + \frac{i}{\lambda} [X^m,X^l] E_{ln},
  \notag
\end{alignat}
where the indices $M,N$ label the target spacetime directions.
$G_{MN}$ and $B_{MN}$ are the background metric and the NS-NS 2-form field respectively.

Now we trace the similar calculations as in ref. \cite{Hya4}.
Setting $X^1 = \cdots = X^6 = 0$, which are parallel to D6-branes,
the Lagrangian for D0-branes evaluated in the background of 
D6-branes becomes as
\begin{alignat}{3}
  \mathcal{L}_{\text{D}0} &= \text{Tr} \bigg( -T_0 \sqrt{f - \frac{f^2}{2\lambda^2}[X^i,X^j]^2
  - f^2 (D_0 X^i)^2 + \frac{f^3}{\lambda^2}(\epsilon_{ijk} \{D_0 X^i, [X^j,X^k]\})^2} 
  \\
  &\qquad\quad - \frac{N}{4r} \frac{X^3}{(X^1)^2 + (X^2)^2} 
  \epsilon_{ij9} \{X^i, D_0 X^j\} \bigg) . \notag
\end{alignat}
Note that the radius $r$ should be understood in terms of $X^i$ as 
$r^2 = \sum (X^i)^2$, and symmetrized trace prescription is replaced with trace
by choosing appropriate orderings. Furthermore we assume that $X^i$ and $A_0$
take following forms
\begin{alignat}{3}
  X^7_{mn} &= \tfrac{1}{2} \rho_{m+1/2} \delta_{m+1,n} 
  + \tfrac{1}{2} \rho_{m-1/2} \delta_{m,n+1} ,& \qquad
  X^9_{mn} &= z_m \delta_{m,n} , \notag
  \\
  X^8_{mn} &= \tfrac{i}{2} \rho_{m+1/2} \delta_{m+1,n} 
  - \tfrac{i}{2} \rho_{m-1/2} \delta_{m,n+1} ,& \qquad
  A_{0\,mn} &= a_m \delta_{m,n}.
\end{alignat}
These forms are useful when we are considering axial asymmetric configurations around $x^9$.
After some calculations we obtain
\begin{alignat}{3}
  \mathcal{L}_{\text{D}0} &= - T_0 \sum_{m=1}^M \sqrt{f + \frac{f^2}{\lambda^2} 
  \big\{ (\rho^2 \delta z^2)_m + (\rho^2 \delta \rho^2)_m - \lambda^2 (\rho^2 \delta a^2)_m \big\} }
  \notag
  \\
  &\quad\,
  - \frac{N}{2} \sum_{m=1}^M \frac{z_m}{\tfrac{1}{2} (\rho_{m+1/2}^2 + \rho_{m-1/2}^2)r_m} 
  (\rho^2 \delta a)_m , 
\end{alignat}
where we defined
\begin{alignat}{3}
  (\rho^2 \delta z^2)_m &= \tfrac{1}{2} ( \rho_{m+1/2}^2 (z_{m+1}-z_m)^2
  + \rho_{m-1/2}^2 (z_m-z_{m-1})^2), \notag
  \\
  (\rho^2 \delta \rho^2)_m &= \tfrac{1}{4} (\rho_{m+1/2}^2 - \rho_{m-1/2}^2)^2,  
  \\
  (\rho^2 \delta a^2)_m &= \tfrac{1}{2} 
  ( \rho_{m+1/2}^2 (a_{m+1}-a_m)^2 + \rho_{m-1/2}^2 (a_m-a_{m-1})^2), \notag
  \\
  (\rho^2 \delta a)_m &= \tfrac{1}{2} (\rho_{m+1/2}^2 (a_{m+1}-a_m) 
  + \rho_{m-1/2}^2 (a_m-a_{m-1}) ). \notag
\end{alignat}
When $M$ is sufficiently large we can take the continuous limit, and the above Lagrangian becomes
\begin{alignat}{3}
  \mathcal{L}_{\text{D}0} &= - T_2 \int dz d\phi~ \sqrt{f \lambda^2 B^2 
  + f^2 \rho^2(1 + {\rho'}^2) - f^2 \rho^2 \lambda^2 E^2 } 
  + \frac{N}{4\pi} \int dz d\phi~ \frac{z}{r} E , 
\end{alignat}
where $E = - \frac{\delta a}{\delta z}$, $B = \frac{1}{\delta z}$ and $\int d\phi = 2\pi$.
This coincides with the action (\ref{eq:D2lag}) after setting $\rho^2 = r^2 - z^2$ and
doing coordinate transformation from $z$ to $\theta$.
Then we trace the same arguments as before, and obtain the gravitational Myers effect.

\section{Fluctuations Around the Spherical D2-D0 Bound State}

In the section \ref{sec:D6}, the dielectric phenomenon in the background of D6-branes
is observed by assuming the spherical configuration of the D2-brane.
Here we examine the stability of this configuration. 
But it is difficult to analyze the stability by adding small fluctuations,
since the energy is given by the elliptic integral.
So in order to make discussions simple, let us consider another distribution of
the magnetic flux on the D2-brane like 
\begin{alignat}{3}
  F_{\theta\phi}=\frac{M}{2} \sin\theta. \label{eq:fake}
\end{alignat}
Of course this configuration is not a classical solution, but the energy
(\ref{eq:D2ene}) of the D2-D0 bound state is simply given by
\begin{alignat}{3}
  \mathcal{H} &= 2\pi^2 \sqrt{ \big(T_2 fr^2\big)^2 
  + f \Big( \frac{T_0 M}{4\pi} \Big)^2}. \label{eq:fakeene}
\end{alignat}
The original solution can be obtained by adding fluctuations around the above configuration,
and it will be possible to understand the stability of the original one.

Now we reexamine the energy (\ref{eq:fakeene}) by adding a fluctuation 
$\bar{r} \epsilon(\theta)$ around $r = \bar{r}$, where $\bar{r}$ gives the minimum of 
the energy. Other fluctuations on $\phi$ and $x^i$ are neglected for simplicity.
By setting $r = \bar{r} (1 + \epsilon)$ and $x^i=0$,
the pullback metric on the D2-brane is given by
\begin{alignat}{3}
  P[G]_{ab} = 
  \begin{pmatrix}
    -f^{-\frac{1}{2}} & 0 & 0 
    \\
    0 & f^{\frac{1}{2}} r^2 + f^{\frac{1}{2}} \bar{r}^2 \epsilon'^2 & 0
    \\
    0 & 0 & f^{\frac{1}{2}} r^2\sin^2\theta
  \end{pmatrix}, \label{eq:mflu}
\end{alignat}
where $a,b$ label $t,\theta,\phi$, and $\epsilon'$ is an abbreviation for $\partial_\theta \epsilon$.
Note that the function $f$ depends on the angular
directions, but $fr = \frac{N\ell_s g_s}{2}$ is just the constant.

Furthermore we add fluctuation of the gauge field $a_\phi(\theta)$ around the background 
of $A_\phi$. Then the component of the gauge field strengths on the D2-brane is described by
\begin{alignat}{3}
  F_{\theta\phi}=\frac{M}{2} \sin\theta \,(1 + \delta). \label{eq:gflu}
\end{alignat}
The fluctuation $\frac{M}{2} \sin\theta \, \delta$ is defined as 
$\frac{M}{2} \sin\theta \, \delta = \partial_\theta a_\phi$
and should satisfy the constraint
\begin{alignat}{3}
  \frac{1}{2\pi} \int d\theta d\phi \frac{M}{2} \sin\theta \,\delta = 0, \label{eq:con}
\end{alignat}
which follows from the flux quantization condition
$\frac{1}{2\pi} \int d\theta d\phi F_{\theta\phi} = M$.

By substituting the pullback metric (\ref{eq:mflu}) and 
the gauge field strength (\ref{eq:gflu}) into the arguments in sec.~\ref{sec:D6},
we obtain the energy for the D2-D0 bound state of the form,
\begin{alignat}{3}
  \mathcal{H} 
  &= \frac{1}{2} \int d\theta 
  \Big[ (4\pi T_2 f r^2)^2 + f (M T_0)^2 ( 1 \!+\! \delta )^2
  + (4\pi T_2 f r)^2 \bar{r}^2 \epsilon'^2 \Big]^{\frac{1}{2}} \notag
  \\
  &= \frac{\bar{\mathcal{H}}}{\pi} \int d\theta 
  \Big[ \frac{1}{3}(1+\epsilon)^2 + \frac{2}{3}\frac{(1+\delta)^2}{1+\epsilon}
  + \frac{1}{3} {\epsilon'}^2 \Big]^{\frac{1}{2}}, \label{eq:D2pot}
\end{alignat}
where $\bar{\mathcal{H}} = \mathcal{H}(\bar{r})$. The last line is 
derived by using the relations $4\pi T_2 f \bar{r}^2 = 2^{-1/2} 
f^{1/2} M T_0 = \frac{2}{\sqrt{3}\pi} \bar{\mathcal{H}}$.
Note that the energy is expanded as
\begin{alignat}{3}
  \mathcal{H} &\sim \bar{\mathcal{H}} + \frac{2 \bar{\mathcal{H}}}{3\pi} 
  \int d\theta \delta. \label{eq:Vflu}
\end{alignat}
The linear term on $\delta$ represents the instability of the configuration (\ref{eq:fake}).
If $\delta$ is a fluctuation around the original configuration (\ref{eq:D2mag}),
the linear term vanishes after imposing the magnetic flux conservation.
But the original configuration is unstable against $\epsilon$ fluctuation.

The final state of this D2-D0 bound state after the perturbation is conjectured as the 
case where $M$ D0-branes are radiated away from $N$ D6-branes.
This configuration has the same quantum numbers as those of the spherical
D2-D0 bound state, and the potential energy at $r=\infty$ is 
zero\footnote{The constant part of $f$ is neglected.}.
During this process the D2-brane will be deformed and decay through the tachyon 
condensation process. The numerical result on this deformation is 
shown in the appendix A.

To avoid the above instability, we just prepare more D6-branes around the original ones properly.
D0-branes will meet the potential barrier before they go away from the original D6-branes.
Then the dielectric D2-D0 configuration, which will depend on the locations of the D6-brnaes, 
becomes stable.

\section{Conclusions and Discussions}

In this paper we observed the gravitational dielectric effect and Myers effect
in the type IIA superstring theory.
In the background of Kaluza-Klein monopoles, the spherical 
D2-brane with the momentum along the compactified circle becomes the classical solution.
This phenomenon is occurred because the tension of the spherical D2-brane balances with
the repulsive force between the Kaluza-Klein monopoles and the momentum along the circle.

As the Kaluza-Klein monopoles are related to the D6-branes via T-S-T duality,
it is expected that the gravitational dielectric effect occurs in the background of
D6-branes. In fact we find that the spherical D2-brane with magnetic flux on it 
becomes the classical solution with the finite radius. Furthermore the similar effect is obtained 
from the viewpoint of D0-branes, that is, D0-branes compose the fuzzy sphere 
in the background of D6-branes. We called this phenomenon the gravitational Myers effect.
In the continuous limit, the gravitational Myers
effect coincides with the gravitational dielectric effect.
The essence of both effects is the fact that the tension of the spherical D2-brane
balances with the repulsive force between the D6-branes and the D0-branes.

It is worth emphasizing that the gravitational dielectric effect and
Myers effect are insensitive to the charges carried by Kaluza-Klein monopoles or D-branes.
Consequently, for instance, a bound state of an anti-D2-brane and D0-branes 
spherically expands in the background of D6-branes or anti-D6-branes.
This is in contrast with the dielectric phenomena induced by the background flux.

By examining the fluctuations around the spherical D2-D0 bound state 
we see that this configuration is unstable
and the final state is conjectured by employing the numerical analyses.
To avoid this instability, we need more D6-branes around the original ones.
Then the dielectric D2-D0 bound state will be stable aginst collapse.

Let us consider generalization of the gravitational dielectric 
effect to other D$p$-branes background. 
In the case of D6-branes background, the gravitational dielectric effect
is induced via the factor $f = f^{\frac{6-3}{2}-\frac{1}{2}+\frac{0}{2}}$ 
in front of $(MT_0)^2$ in the potential energy (\ref{eq:pot}).
In cases of a spherical D$q$-brane with magnetic flux on it 
in the background of D$p$-branes, this factor is modified as
$f^{\frac{p-3}{2}-\frac{1}{2}+\frac{q-2}{2}}$. 
Here the D$q$-brane is completely transverse to the D$p$-branes.
The first part of the exponent comes from the contribution of the dilaton field 
and the second one does from that of $-G_{00}$.
The third term appears from $\prod_{11-q}^9 G_{ii}$, where
$(q-2)$ is the number of world-volume coordinates which are normal to 
both the background D$p$-branes and the sphere.
From this we see that the dielectric effect occurs if $p+q=8$.
These are all obtained by T-duality transformations of the spherical D2-D0 bound state 
in the background of D6-branes.

In this paper we focused on the single D2-D0 bound state wrapped around
D6-branes. But this system can be generalized to the system
where multiple D2-D0 bound states are wrapped around each bunch of D6-branes.
Here the bunches of D6-branes are separated each other.
In case some of them are sufficiently close, the spherical D2-D0 bound states which are 
wrapped around them will make a transition to a single D2-D0 bound state. 
The configuration of the D2-brane or the fuzzy D0-branes will depend on the 
locations of the bunches of D6-branes, and it is interesting to analyze
the dynamics from the viewpoint of tachyon condensation\cite{Hya,Has,Kim}.
After the tachyon condensation, the D0-branes will compose a general fuzzy surface
and representations for the matrices $X^i$ will be described as in ref. \cite{Shi}.

\section*{Acknowledgements}

I would like to thank members of Komaba particle theory group,
especially Koji Hashimoto and Hidehiko Shimada 
for useful discussions. The work was supported in part by 
the Grant-in-Aid for JSPS fellows.

\appendix
\section{Numerical Results on the Deformation of Spherical D2-D0 Bound State}

Numerical calculations about the potential energy (\ref{eq:D2pot}) are shown
in this appendix. 
Fluctuations $\epsilon$ and $\delta$ are expanded by Legendre polynomials as
\begin{alignat}{3}
  \epsilon(x,t) &= \sum_{l=0}^{15} \epsilon_l(t) P_l(x), \quad
  \delta(x,t) &= \sum_{l=1}^{15} \delta_l(t) P_l(x).
\end{alignat}
The expansions are limited to $l=15$, which is purely numerical reason.
The deformation of the D2-D0 bound state in the background of D6-branes is 
parametrized by $t$, and $\epsilon_l(0) = \delta_l(0) = 0$ for all $l$.

\begin{figure}[h]
\begin{center}
  \includegraphics[width=7cm,height=7cm,keepaspectratio]{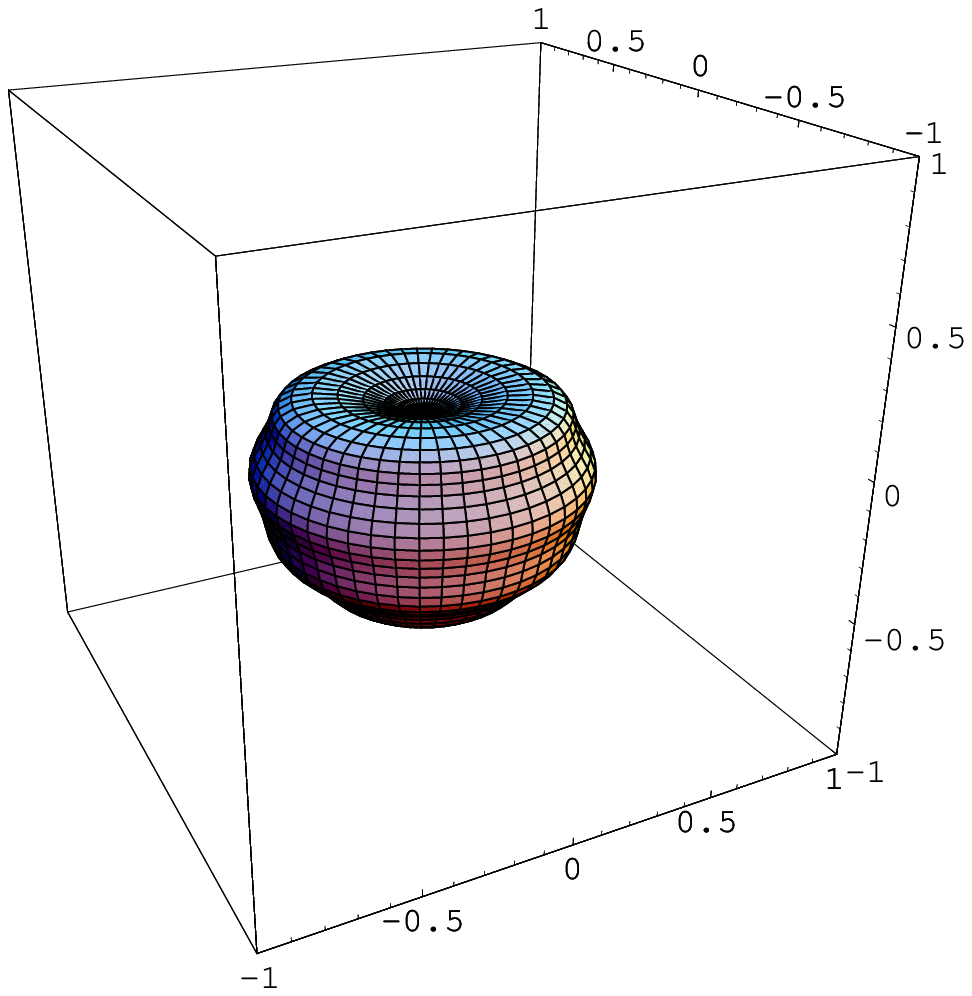}
  \hspace{1cm}
  \includegraphics[width=7cm,height=7cm,keepaspectratio]{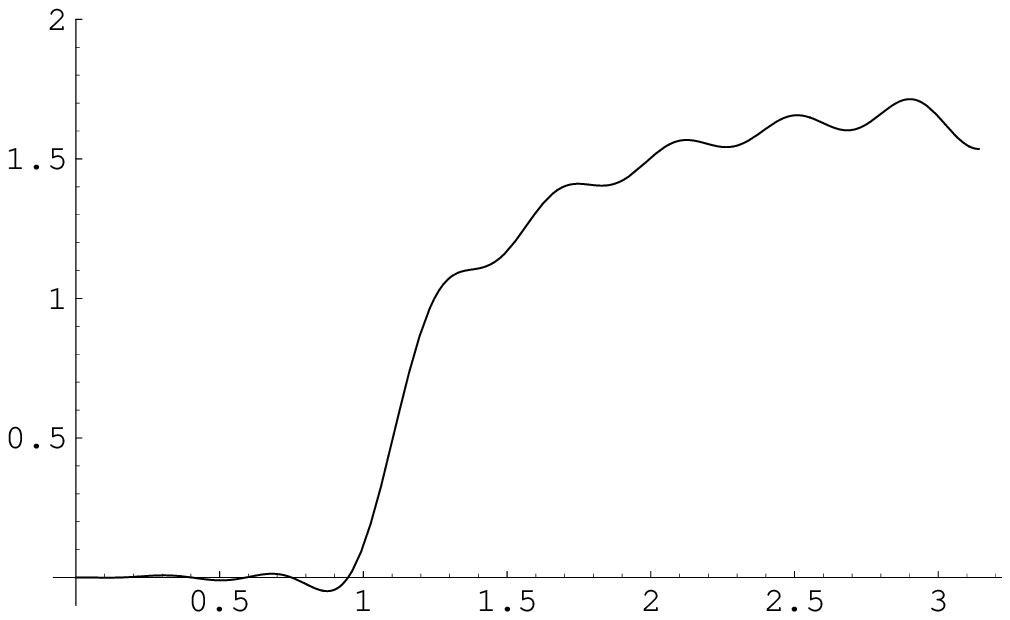}
\begin{picture}(300,0)
  \put(20,0){$(a)$}
  \put(272,0){$(b)$}
  \put(370,20){$\theta$}
  \put(170,142){$1+\delta$}
\end{picture}
  \caption{(a) The figure of D2-D0 bound state at $t=1$. (b) The plot of $(1+\delta)$
  at $t=1$.}
  \label{fig:D01}
\end{center}
\end{figure}

\begin{figure}[h]
\begin{center}
  \includegraphics[width=7cm,height=7cm,keepaspectratio]{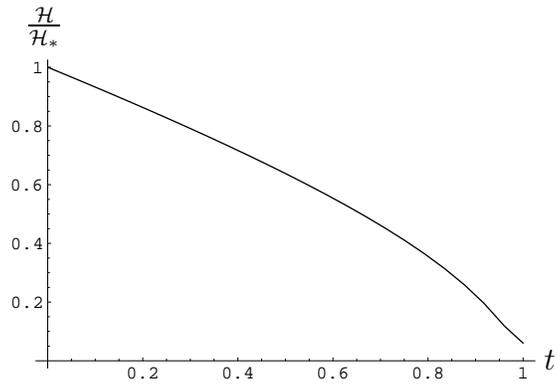}
\begin{picture}(300,0)
  \put(57,147){$\frac{\mathcal{H}}{\mathcal{H}_\ast}$}
  \put(253,20){$t$}
\end{picture}
  \caption{The potential energy along the deformation (\ref{eq:D0}). 
  The final state will be the case where $M$ D0-branes are scattered from $N$ D6-branes.}
  \label{fig:D03}
\end{center}
\end{figure}

We consider the following deformation:
\begin{alignat}{3}
  (\epsilon_0(t),\cdots, \epsilon_{15}(t))
  &= t (-0.59,\, -0.39,\, -0.11,\, 0.079,\, 0.11,\, 0.030,\, -0.052,\, -0.060, \notag
  \\
  &\; -0.0051,\, 0.043,\, 0.034,\, -0.014,\, -0.038,\, -0.0078,\, 0.033,\, -0.0059), \notag
  \\
  (\delta_1(t),\cdots, \delta_{15}(t)) 
  &= t (-1.00,\, -0.38,\, 0.13,\, 0.31,\, 0.18,\, -0.060,\, -0.19,\, -0.13, \notag
  \\
  &\quad 0.034,\, 0.15,\, 0.11,\, -0.030,\, -0.14,\, -0.091,\, 0.12) \label{eq:D0}
\end{alignat}
The figures of $((1+\epsilon)\sin\theta\cos\phi, (1+\epsilon)\sin\theta\sin\phi, 
(1+\epsilon)\cos\theta)$ and $1+\delta$ at $t=1$ are drawn in fig. \ref{fig:D01}. 
The potential energy along this deformation is plotted
in fig. \ref{fig:D03}. From this we see that the D2-brane is deformed 
and D0-branes are distributed around the southern hemisphere.
After the tachyon condensation process, the D2-brane will vanish and D0-branes
will be scattered to the infinity.

\newpage

\newpage
\begin{thebibliography}{99}



\bibitem{Mye1} See for example, 
R. C. Myers,
\textit{``Nonabelian Phenomena on D Branes''}, 
Class. Quant. Grav. {\bf 20} (2003) S347; hep-th/0303072.

\bibitem{BDS} C. Bachas, M. R. Douglas and C. Schweigert, 
\textit{``Flux Stabilization of D-branes''}, 
JHEP {\bf 0005} (2000) {\bf 048}; hep-th/0003037.

\bibitem{Hya2} Y. Hyakutake, 
\textit{``Expanded Strings in the Background of NS5-branes via a M2-brane,
a D2-brane and D0-branes''}, 
JHEP {\bf 0201} (2002) 021; hep-th/0112073.

\bibitem{Hya3} Y. Hyakutake, 
\textit{``Expanded Strings in the Background of NS5-branes''}, 
Grav. Cosmol. {\bf 9} (2003) 37.

\bibitem{MST} J. McGreevy, L. Susskind and N. Toumbas, 
\textit{``Invasion of the Giant Gravitons from Anti-de Sitter Space''}, 
JHEP {\bf 0006} (2000) 008; hep-th/0003075.

\bibitem{DJM} S. R. Das, A. Jevicki and S. D. Mathur, 
\textit{``Giant Gravitons, BPS bounds and Noncommutativity''}, 
Phys. Rev. {\bf D63} (2001) 044001; hep-th/0008088.

\bibitem{Asa} M. Asano,
\textit{``Noncommutative Branes in D-brane Backgrounds''}, \\
Int. J. Mod. Phys. {\bf A17} (2002) 4733; hep-th/0106253.

\bibitem{CHC} M. Costa, C. Herdeiro, L. Cornalba,
\textit{``Fluxbranes and the Dielectric Effect in String Theory''}, 
Nucl. Phys. \textbf{B619} (2001) 155; hep-th/0105023.

\bibitem{ML} M. Li, 
\textit{``Boundary States of D-Branes and Dy-Strings''}, 
Nucl. Phys. \textbf{B460} (1996) 351; hep-th/9510161.

\bibitem{Dou} M. R. Douglas, 
\textit{``Branes within Branes''}, hep-th/9512077.

\bibitem{ARS} A. Y. Alekseev, A. Recknagel and V. Schomerus,
\textit{`` Noncommutative World Volume Geometries: 
Branes on SU(2) and Fuzzy Spheres''},
JHEP {\bf 9909} (1999) 023; hep-th/9908040. 

\bibitem{ARS2} A. Y. Alekseev, A. Recknagel and V. Schomerus,
\textit{`` Brane Dynamics in Background Fluxes and Noncommutative Geometry''},
JHEP {\bf 0005} (2000) 010; hep-th/0003187.

\bibitem{HNS} Y. Hikida, M. Nozaki and Y. Sugawara, 
\textit{``Formation of Spherical D2 Brane from Multiple D0 Branes''}, 
Nucl. Phys. {\bf B617} (2001) 117; hep-th/0101211.

\bibitem{Mye2} R. C. Myers,
\textit{``Dielectric Branes''}, JHEP {\bf 9912} (1999) 022; hep-th/9910053.

\bibitem{GP} D. J. Gross and M. J. Perry,
\textit{``Magnetic Monopoles in Kaluza-Klein Theories''}, 
Nucl. Phys. {\bf B226} (1983) 29.

\bibitem{OV} H. Ooguri and C. Vafa,
\textit{``Two-dimensional Black Hole and Singularities of CY Manifolds''}, 
Nucl. Phys. {\bf B463} (1996) 55.

\bibitem{BG} J. de Boer, E. Gimon, K. Schalm and J. Wijnhout,
\textit{``Evidences for a Gravitational Myers Effect''},
hep-th/0212250.

\bibitem{Tow} P. K. Townsend, 
\textit{``The Eleven-dimensional Supermembrane Revisited''}, \\
Phys. Lett. {\bf B350} (1995) 184; hep-th/9501068.

\bibitem{Lei} R. G. Leigh, 
\textit{``Dirac-Born-Infeld Action From Dirichlet Sigma Model''}, \\
Mod. Phys. Lett. {\bf A4} (1989) 2767.

\bibitem{GM} G. W. Gibbons and Kei-ichi Maeda, 
\textit{``Black Holes and Membranes in Higher Dimensional Theories with Dilaton Fields''},
Nucl. Phys. {\bf B298} (1988) 741.

\bibitem{Wit} E. Witten, 
\textit{``Bound States Of Strings And $p$-Branes''}, 
Nucl. Phys. \textbf{B460} (1996) 335; hep-th/9510135.

\bibitem{Tse} A. A. Tseytlin, 
\textit{``Born-Infeld Action, Supersymmetry and String Theory''}, \\
hep-th/9908105.

\bibitem{Tse2} A. A. Tseytlin,
\textit{``On Nonabelian Generalization of Born-Infeld Action 
in String Theory''}, Nucl. Phys. {\bf B501} (1997) 41; hep-th/9701125.

\bibitem{TR} W. Taylor, M. V. Raamsdonk, 
\textit{``Multiple Dp-branes in Weak Background Fields''}, \\
Nucl. Phys. \textbf{B573} (2000) 703; hep-th/9910052.

\bibitem{Hya4} Y. Hyakutake, 
\textit{``Notes on the Construction of the D2-brane from Multiple D0-branes''},
Nucl. Phys. {\bf B675} (2003) 241; hep-th/0302190.

\bibitem{Hya5} Y. Hyakutake, 
\textit{``Fuzzy BIon''},
Phys. Rev. {\bf D68} (2003) 046003; hep-th/0305019.

\bibitem{Hya} Y. Hyakutake, 
\textit{``Torus-like Dielectric D2-brane''}, \\
JHEP {\bf 0105} (2001) 013; hep-th/0103146.

\bibitem{Has} K. Hashimoto, 
\textit{``Dynamical Decay of Brane Anti-Brane and Delectric Brane''}, \\
JHEP {\bf 0207} (2002) 035; hep-th/0204203.

\bibitem{Kim} Y. Kimura, 
\textit{``Myers Effect and Tachyon Condensation''}, hep-th/0309082.

\bibitem{Shi} H. Shimada, 
\textit{``Membrane Topology and Matrix Regularization''}, hep-th/0307058.



\end{thebibliography}
\end{document}